\documentclass[ 3p,10pt, twocolumn]{elsarticle}

\usepackage{float}
\usepackage{graphicx}
\usepackage[pdfversion=1.4]{hyperref}

\usepackage{amssymb}
\usepackage{amsmath}
\usepackage{listings}
\usepackage{color}
\definecolor{cadmiumgreen}{rgb}{0.0, 0.42, 0.24}
\definecolor{deepcarrotorange}{rgb}{0.91, 0.41, 0.17}
\definecolor{darkpowderblue}{rgb}{0.0, 0.2, 0.6}\definecolor{blue(ryb)}{rgb}{0.01, 0.28, 1.0}
\biboptions{sort&compress,numbers,square,comma}

\newcounter{bla}

\journal{Computer Physics Communications}

\newcommand{\e}{\mathrm{e}}

\begin{document}

\begin{frontmatter}



\title{\textsc{Bill2d} -- a software package for classical two-dimensional Hamiltonian systems}


\author[a]{J. Solanp\"a\"a\corref{author}}
\cortext[author] {Corresponding author}
\author[b]{P. J. J. Luukko}
\author[a]{E. R\"as\"anen}
\address[a]{Department of Physics, Tampere University of Technology, Tampere FI-33101, Finland}
\address[b]{Nanoscience Center, Department of Physics, University of Jyv\"askyl\"a, FI-40014 Jyv\"askyl\"a, Finland}

\begin{abstract}
We present \textsc{Bill2d}, a modern and efficient C$++$ package for classical simulations of two-dimensional
Hamiltonian systems. \textsc{Bill2d} can be used for various billiard and diffusion problems with
one or more charged particles with interactions, different external potentials, an external magnetic field, periodic
and open boundaries, etc. The software package can also calculate many key quantities in complex systems
such as Poincar\'e sections, survival probabilities, and diffusion coefficients.
While aiming at a large class of applicable systems, the code also strives for ease-of-use, efficiency, and modularity
for the implementation of additional features. The package comes along with a user guide, a developer's manual, and a
documentation of the application program interface (API).
\end{abstract}

\begin{keyword}
Classical mechanics; billiards; nonlinear dynamics; chaos; transport; diffusion; numerical simulations; molecular dynamics\\
\vspace*{1cm}
This is the final preprint of the article published in \href{http://dx.doi.org/10.1016/j.cpc.2015.10.006}{Comp.  Phys. Comm. \textbf{199}, 133-138 (2016)}.
\end{keyword}

\end{frontmatter}

\clearpage
{\noindent\bf PROGRAM SUMMARY}\\
\begin{small}
\noindent
{\em Program Title:}  Bill2d  \\
{\em Journal Reference:}                                      \\
{\em Catalogue identifier:}                                   \\
{\em Licensing provisions:}  GNU General Public License version 3   \\
{\em Programming language:}  C$++$(14)  \\
{\em Computer:}  Tested on x86 and x86\_64 architectures.    \\
{\em Operating system:}  Tested on Linux, and OS X versions 10.9-10.11.                                      \\
{\em Compilers:} C$++$14 compliant compiler. Tested compilers: Clang 3.7, GCC 4.9-5.2, and Intel C$++$ compiler 15. \\
{\em RAM:} Simulation dependent: kilobytes to gigabytes                                          \\
{\em Parallelization:} Shared memory parallelization when simulating ensembles of systems.                             
{\em Classification:}
4.3 Differential equations, 7.8 Structure and Lattice dynamics, 7.9 Transport properties, 7.10 Collisions in solids,
16.9 Classical methods\\
{\em External routines/libraries:}                            \\
Boost, CMake, GSL, HDF5; and optionally GoogleMock, GoogleTest, and Doxygen\\
{\em Nature of problem:}\\
  Numerical propagation of classical two-dimensional single and many-body systems, possibly in a magnetic field, and calculation of relevant quantities such as Poincar\'e sections, survival probabilities, diffusion coefficients, etc.
   \\
{\em Solution method:}\\
  Symplectic numerical integration of Hamilton's equations of motion in Cartesian coordinates, or solution of Newton's equations of motion if in a magnetic field. The program implements several well-established algorithms.
   \\
{\em Restrictions:}\\
  Pointlike particles with equal masses and charges, although the latter restrictions are easy to lift.
   \\
{\em Unusual features:}\\
Program is efficient, extremely modular and easy to extend, and allows arbitrary particle-particle interactions.   \\
{\em Additional comments:}\\
The source code is also available at \url{https://bitbucket.org/solanpaa/bill2d}. See \path{README} for locations of user guide, developer manual, and API docs.
   \\
{\em Running time:}\\
  From milliseconds to days, depends on type of simulation.
\end{small}

\section{Introduction}
Hamiltonian systems still offer a wide range of yet unexplored territories and interests for the study of nonlinear dynamics and chaos.
This is evident from the recent progress, e.g., in the study of transmission, escape rates, and survival probabilities of open systems and recurrences in closed systems ~\cite{chaos_alive,circlebilliards,solanpaa_msc,openspherical,stadium_transport,universal_decays,bouncertransport,round_border_escape,escape_dettmann,openmushroom}, stickiness and marginally unstable periodic orbits~\cite{openmushroom,jpn_stickiness,manyfaces,fine_stickiness}, and
detailed analysis of the graphene-like Lorentz gas and related systems~\cite{lorentzreview,flower_lorentz,3dlorentz,recent_hard_wall_lorentz,magnetoresistance,magnetoresistance2,diffusion_repulsive_attractive,diffusion_2d_soft},
and this list barely scratches the surface. As pointed out by Bunimovich and Vela-Arevalo, ``Chaos theory is very much alive''~\cite{chaos_alive}.

Two-dimensional (2D) Hamiltonian systems (i.e., with two coordinate
dimensions), such as dynamical billiards, have the advantage of being simple,
but not too simple. That is, they are easy to study and visualize, but even a
single-particle system can display a large variety of complex phenomena of Hamiltonian chaos.
Many of these systems can
also be realized as semiconductor nanostructures
\cite{nakamura,micolich}. While governed by quantum
mechanics, these nanostructures have ballistic regions where a classical
treatment applies to some extent \cite{stadium_transport}.

2D billiards (including soft potentials) with point-like particles can be simulated
either using the exact solution, the exact Poincar\'e map,
or by solving Hamilton's or Newton's equations of motion.
The exact solution is often unknown, apart from single-particle systems with simple geometries.
Therefore, one often resorts to numerical solution of the classical equations of motion.

In this paper we introduce \textsc{Bill2d}, a code for generic, classical 2D systems with a focus on
the study of chaos and nonlinear dynamics. The code can handle single or many particles,
hard-wall boundaries (as in traditional billiards), particle-particle interactions,
external potentials, periodic systems, a magnetic field, etc.
Written in modern C$++$, \textsc{Bill2d} is modular and easy to extend without sacrificing
speed or ease-of-use.

The paper is organized as follows: In Sec. \ref{sec:systems} we describe
the class of systems to which \textsc{Bill2d} can be applied.
In Sec. \ref{sec:algorithms} we briefly discuss the numerical propagation algorithms
and in Sec.~\ref{sec:implementation} implementation and structure of the code.
In Sec.~\ref{sec:examples} we give a few numerical examples of what can be simulated with \textsc{Bill2d}.
In Sec. Sec.~\ref{sec:summary} we finish with a brief summary of the paper.

\section{Systems}\label{sec:systems}
\subsection{Overview}
\textsc{Bill2d} is designed for classical dynamics of interacting particles in two-dimensions.
The program can deal with both traditional billiards with hard-wall boundaries as well as with
soft external potentials including also periodic systems. Single- and multiparticle systems with a generic form for
the particle-particle interaction can be simulated, and an external magnetic field can
be included. Transport calculations are supported by allowing the creation and removal of particles
during the simulation.

When treating systems \emph{without} a magnetic field, we propagate the particles via Hamilton's equations of motion using symplectic algorithms. Systems \emph{with} a magnetic field are propagated using Newton's equations of motion.

The program is written using Cartesian coordinates and atomic units, i.e., we set masses, charges, and the Coulomb constant to unity.
All the particles are point-like and with equal masses and charges (as, e.g., electrons in semiconductors),
although the equal mass and charge requirement can be lifted with minor modifications to the code.

To clarify, systems without an external magnetic field are described by the $N$-particle Hamiltonian ($N\geq 1$)
\begin{equation}\label{eqn:hamiltonian}
H=\sum\limits_{i=1}^N \left(\frac{1}{2} \Vert\mathbf{p}_i\Vert^2 + V(\mathbf r_i) \right)+ \sum\limits_{i<j} W(\mathbf r_i, \mathbf r_j),
\end{equation}
where $\mathbf r_i$ and $\mathbf p_i$ are the position and momentum of the $i$th particle,  $V(\mathbf r)$ is the total external single-particle potential, and $W(\mathbf r_i, \mathbf r_j)$ the interparticle potential. In principle, potentials $V$ and $W$ can be arbitrary, but in the case of attracting singularities, new propagation algorithms should be implemented.

When the system includes a magnetic field, we use the Newtonian formulation instead of Hamiltonian formulation.
The equations of motion are
\begin{equation}\label{eqn:newton}
\begin{split}
\frac{{\rm d}^2\mathbf r_i}{{\rm d}t^2} = &-\nabla_{\mathbf r_i} V(\mathbf r_i)-\sum\limits_{j\neq i} \nabla_{\mathbf r_i} W(\mathbf r_i, \mathbf r_j) \\
&+ B\left(\frac{{\rm d} r_{i,y}}{{\rm d}t}\mathbf e_x - \frac{{\rm d} r_{i,x}}{{\rm d}t} \mathbf e_y \right){},
\end{split}
\end{equation}
where $r_{i,x/y}$ is the $x$ or $y$  coordinate of the position of the $i$th particle and $\mathbf e_{x/y}$ the corresponding unit vector. The first term in Eq. \eqref{eqn:newton} is the force due to the external potential, the second term the interparticle interaction, and the third term the Lorentz force with the magnetic flux density $B$ perpendicular to the two-dimensional plane.

The systems can also include hard-wall billiard boundaries. The (fully elastic) collision of the $i$th particle with the billiard boundary is described by the transformation
\begin{equation}
\begin{aligned}
\label{eqn:billiard_reflection_law}
\mathbf{r}_i &\to\mathbf{r}_i\\
\mathbf{p}_i &\to \mathbf{p}_i-2 \left[\mathbf n (\mathbf{r}_i)\cdot \mathbf{p}_i\right]\, \mathbf{n}(\mathbf{r}_i),
\end{aligned}
\end{equation}
where $\mathbf n(\mathbf r_i)$ is a unit vector perpendicular to the corresponding boundary at position $\mathbf r_i$.

\subsection{Note on Coulomb interaction and the interaction strength parameter}
When describing Coulomb-interacting systems,
we introduce the \emph{interaction strength} parameter $\alpha$.
Essentially the Coulomb-potential is given by
\begin{equation}
W(\mathbf r_i, \mathbf r_j) = \sum_{i<j} \frac{\alpha}{\Vert \mathbf r_i-\mathbf r_j \Vert}.
\end{equation}
By fixing the total energy and length scale (i.e., the size of the billiard table) of the system, we can actually study all length and energy scales of all geometrically similar systems just by varying $\alpha$.

Naturally this trick works only if the system does not have external potentials (apart from a hard-wall boundary) or a magnetic field. In case of a magnetic field
and/or external potentials, also those have to be adjusted. For detailed derivation of the corresponding scale transformations and demonstration of the idea, we refer the reader to Appendix B of Ref.~\cite{solanpaa_msc}.

\section{Algorithms}\label{sec:algorithms}
\subsection{Propagation without a magnetic field}
The propagator, for Hamiltonian systems with the Hamiltonian $H$, is
\begin{equation}
\exp(t \left\{\cdot, H\right\}),
\end{equation}
where $\left\{\cdot, \cdot\right\}$ is the Poisson bracket operator.

The systems we are interested in are described by Hamiltonians [Eq. \eqref{eqn:hamiltonian}] that can be split to two parts as $H=K(\mathbf p)+U(\mathbf r)$, where all the momentum dependence is in $K$, and all the coordinate dependence is in $U$. As the Hamiltonian is separable, there are plentiful of \emph{explicit} algorithms for the time propagation. This is in contrast to general Hamiltonian systems where the symplectic integrators are typically implicit and hence computationally more demanding.

The algorithms we have implemented are based on \emph{split operator schemes},
where the propagator, $\exp\left(\Delta t \left\{\cdot,H\right\}\right)$ is approximated by some product of $\exp\left(const \cdot\Delta t \left\{\cdot,K\right\}\right)$ and $\exp\left(const\cdot\Delta t \left\{\cdot,U\right\}\right)$. The operation of these latter exponentials can be carried out exactly -- to numerical accuracy -- allowing for an easy implementation of split operator schemes.
For example, the velocity Verlet scheme~\cite{alg_paper1} can be expressed as
\begin{equation}
\e^{\Delta t \left\{\cdot,H\right\}} \approx \e^{\frac{1}{2}\Delta t \left\{\cdot,U\right\}} \e^{\Delta t \left\{\cdot,K\right\}} \e^{\frac{1}{2}\Delta t \left\{\cdot,U\right\}}.
\end{equation}

We have implemented symplectic algorithms of orders 2--6 for systems without a magnetic field. The algorithms (from Refs.~\cite{alg_paper1,alg_paper2,alg_paper3,alg_paper4}) are specified in \path{USERGUIDE}.

\subsection{Propagation in a magnetic field}
In a magnetic field, the systems are propagated via Newton's equations of motion [Eq. \eqref{eqn:newton}] as we are unaware of any efficient symplectic algorithms for this class of systems.
We have implemented two algorithms: (i) the second order Taylor expansion scheme developed by Spreiter and Walter~\cite{spreiter-walter} and (ii)
the fourth order algorithm developed by Y. He \emph{et} al. in Ref. \cite{4th_order_magfield_algorithm}.

The algorithm by Spreiter and Walter has been later shown to be an energy
conserving split-operator scheme~\cite{pre_77_066401}, and the fourth order algorithm
is volume preserving (and also preserves the total energy extremely well)~\cite{4th_order_magfield_algorithm}.
These schemes explicitly incorporate the effects of the magnetic field into the propagation formulas,
and allow -- in principle -- an arbitrarily strong magnetic field without the need for a smaller time step.

\section{Implementation}\label{sec:implementation}

\subsection{Overview}
\textsc{Bill2d} is designed as a modular, object-oriented package,
and is written in standards-compliant C$++$ following the \texttt{International Standard ISO/IEC 14882:2014(E) Programming Language C++} (aka C$++$14)~\cite{c++standard}.
The package contains three binaries: \texttt{bill2d} for single simulation of a (many-body) trajectory, \texttt{bill2d\_escape} for the calculation of escape rates in open billiards, and  \texttt{bill2d\_diffusion\_coefficient} for the calculation of diffusion coefficients in periodic systems. All the binaries obey similar inputs; more details are given in \path{USERGUIDE}.

The binaries make use of the \texttt{libbill2d} library, which is created
during the compilation process. Most of the classes and methods of the library are enclosed in the namespace \texttt{bill2d}. This library can be used for easy implementation of additional binaries.
A tutorial for developing new binaries and features can be found in \texttt{docs/developer\_manual.pdf}, which also describes the implementation in detail.

\subsection{Input and initial positions}

The class \texttt{Parser} handles parsing of the input either from command line, configuration file, or a combination of these.
The option parsing is implemented with the help of \texttt{Boost::program\_options} library.

The input defines the system (number of particles, magnetic field, potentials, periodicity, etc.) and simulation parameters (algorithm, time step, simulation time, etc.).
The initial positions and velocities of the particles can either be supplied manually, or \textsc{Bill2d} can randomize them.

The default random initial conditions are calculated as follows: First the initial positions of the particles are randomized either within a given hard-wall billiard table, a unit cell, or a user-supplied rectangular area. Next, the kinetic energy is distributed evenly among the particles, and the directions of the velocities are randomized.
Note also that implementation of new randomization procedures for the initial conditions is easy.

\subsection{Application programming interface}

The package consists of several classes each designed usually for a single purpose: \texttt{Table} handles collisions with hard-wall boundaries, \texttt{Datafile} handles saving of data, etc.
The easiest way for a developer to create instances of all the necessary classes is to use the \texttt{Parser} to get all the input parameters bundled in an instance of \texttt{ParameterList},
which can then be passed on to \texttt{BilliardFactory}. This factory class can be used to create instances of \texttt{Billiard},
which bundles all the other necessary objects together. Furthermore, instances of the \texttt{Billiard} class save data automatically during simulation and upon destruction. In all simplicity, a minimal working example could be as follows.

\noindent\begin{minipage}{\linewidth}
\small
\begin{lstlisting}[frame=single]
#include "bill2d/parser.hpp"
#include "bill2d/billiard_factory.hpp"

using namespace bill2d;

int main( int argc, char *argv[] ) {

   @Parser@ parser( argc, argv );

   auto params =
          parser.getParameterList();

   @BilliardFactory@ factory( params );

   auto billiard = factory.get_bill();

   // Set initial condition
   billiard->prepare(3.0 /* energy */);

   // Get reference to time.
   // This variable is kept up to date.
   const auto& time =
          billiard->get_time();

   while ( time < 5.0 )
          billiard->propagate();

} // Saving data upon
  // destruction of billiard
\end{lstlisting}
\end{minipage}

For more complex examples, see files \path{src/bill2d_single_run.cpp}, \path{src/bill2d_escape.cpp}, and \path{src/bill2d_diffusion_coefficient.cpp}, and the developer manual \path{docs/developer_manual.pdf}

\subsection{Datafile}

The program saves its data (trajectories, energies, etc.) to a HDF5-file. The datafile can be accessed afterwards for analysis with several tools including, e.g., the ready made scripts included in the
\textsc{Bill2d} package, different programming languages (at least Python, C, C$++$, Fortran, and Java) and programs such as Matlab and Mathematica. The \path{USERGUIDE} describes in more detail how to access the HDF5-files.

\subsection{Test suite}

The software is bundled with comprehensive unit tests, which currently cover most of the program. Additional tests can be run with the Python script \path{test/propagator_tests.py},
which requires the user to download reference data from the URL specified in \path{README}.

\section{Numerical examples}\label{sec:examples}
\subsection{Trajectories}
\begin{figure}[H]\centering
\includegraphics[width=0.5\linewidth]{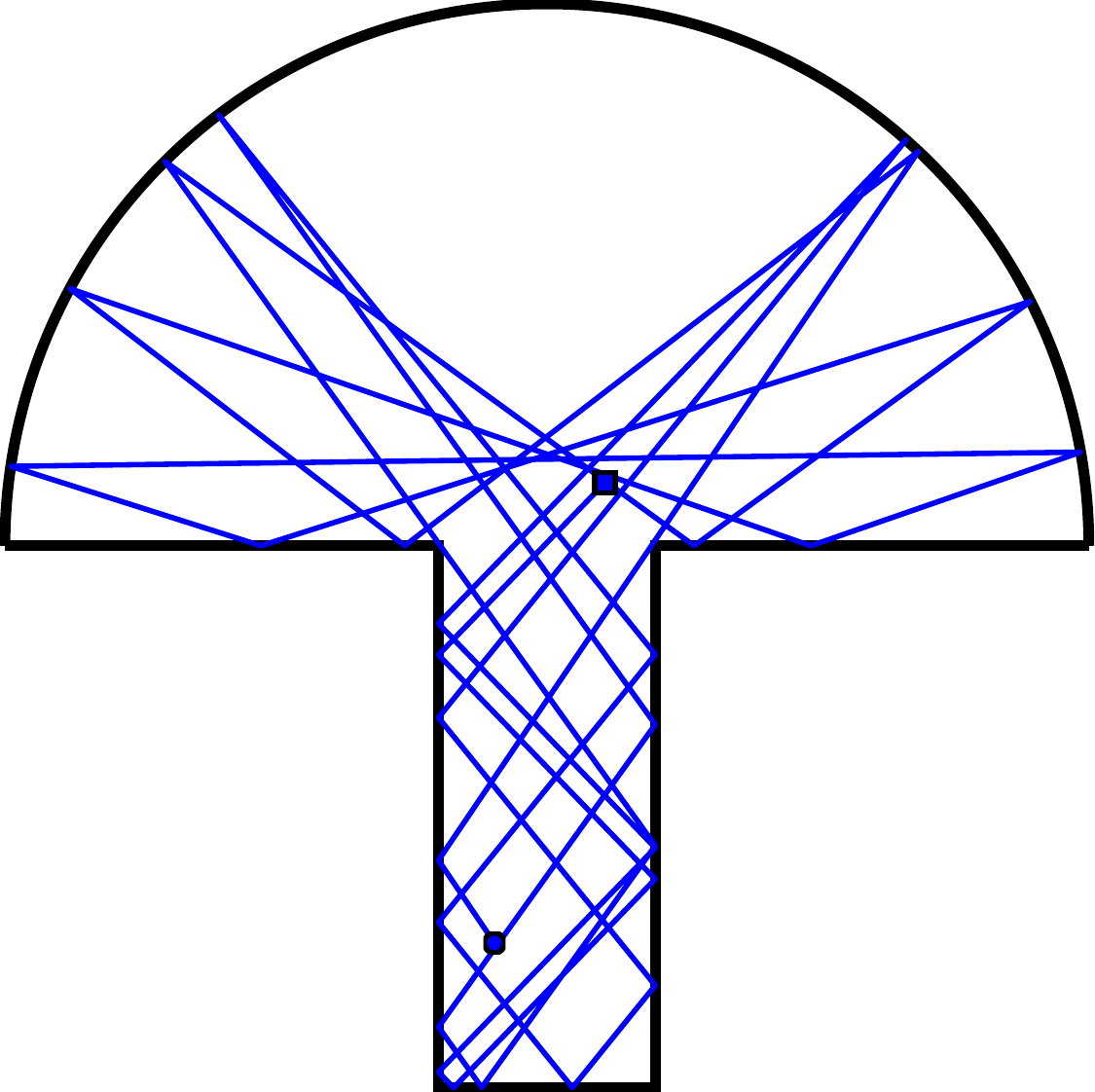}%
\caption{\label{fig:mushroom} Single-particle billiards in a Bunimovich mushroom.}
\end{figure}

A fundamental concept in classical point-particle mechanics is a trajectory, i.e., the curve that the particle draws in the coordinate space.
In Fig.~\ref{fig:mushroom} we show a single-particle trajectory in the Bunimovich mushroom billiards~\cite{mushroom} calculated with the \texttt{bill2d} binary. The trajectory has been drawn with the \texttt{draw\_trajectories} script, which is bundled with the software package.

\begin{figure}[H]\centering
\includegraphics[width=0.5\linewidth]{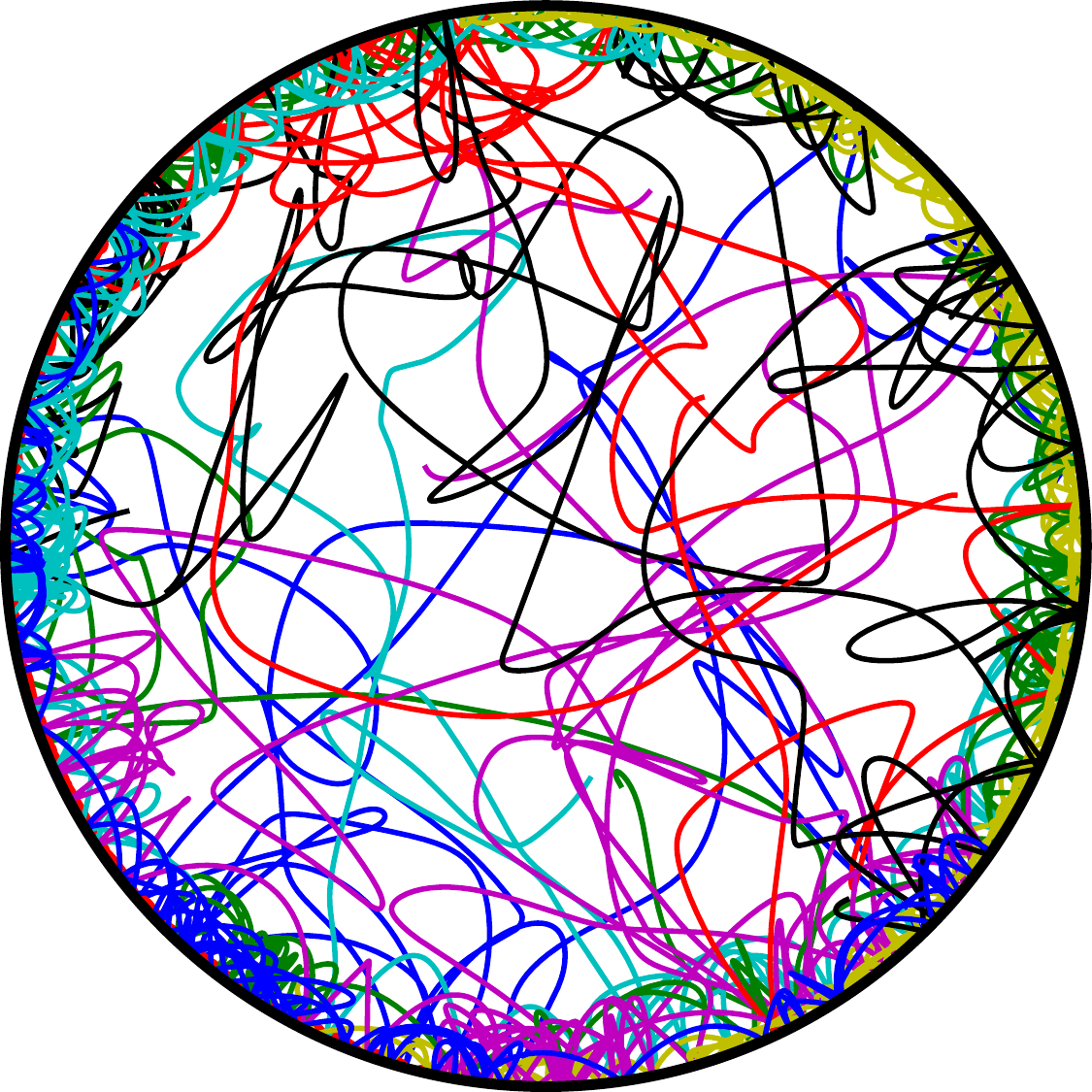}%
\caption{\label{fig:circle} Circular billiards with 15 Coulomb-interacting particles.}
\end{figure}

Similarly, one can consider trajectories of a many-particle system, as in Fig.~\ref{fig:circle}, which shows 15 Coulomb-interacting particles inside a circular billiard table. Here several different trajectories share the same color.


\subsection{Poincar\'e sections}
The phase space can be studied by
selecting a two-dimensional subspace of the full phase space, and visualizing all crossings of a phase space trajectory (or trajectories) with the chosen subspace, which is called the Poincar\'e section.

\begin{figure}[H]\centering
\includegraphics[width=\linewidth]{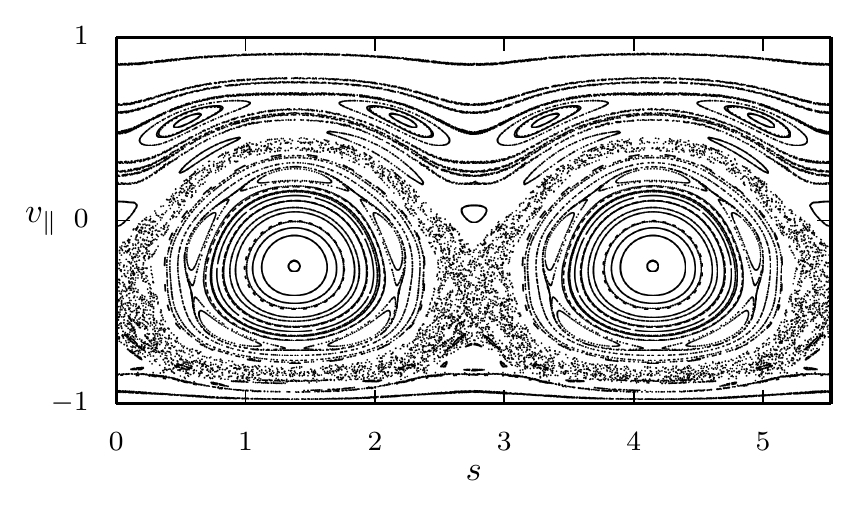}%
\caption{\label{fig:ellipse_bmap} Poincar\'e section of an elliptic billiard with semi-major and -minor axes 1 and $3/4$ in a magnetic field with Larmor radius $r=3$.}
\end{figure}

In Fig.~\ref{fig:ellipse_bmap} we show the Poincar\'e section of a single-particle elliptic billiards in a magnetic field calculated with the \texttt{bill2d} binary; several initial positions are considered in this figure. Here the Poincar\'e section shows the collisions with the billiard table, parametrized by the arc length $s$ and the tangential velocity $v_\parallel$. The Poincar\'e section consists of several elliptic and parabolic KAM islands (after Kolmogorov, Arnold, and Moser~\cite{K,A,M}) that correspond to regular motion, as well as a chaotic sea indicated by irregularly distributed points. This system has previously been studied in, e.g., Ref.~\cite{robnikberrymagfield}.

\subsection{Escape rates in open billiards}
Open billiards have one or many holes in the billiard boundary through which the particle(s) can escape the system.
The object of interest here is the survival probability or the escape rate. The survival probability $P_\mathrm{s}(t)$ gives the probability that the particle has not exited the system before time $t$, and the  escape rate $p_\mathrm{e}(t)$ gives the probability density of escape within a narrow time interval around $t$. These two are naturally related by
\begin{equation}
\frac{\mathrm{d} P_\mathrm{s}(t)}{\mathrm{d}t} = - p_\mathrm{e}(t).
\end{equation}

\begin{figure}[H]\centering
\includegraphics[width=\linewidth]{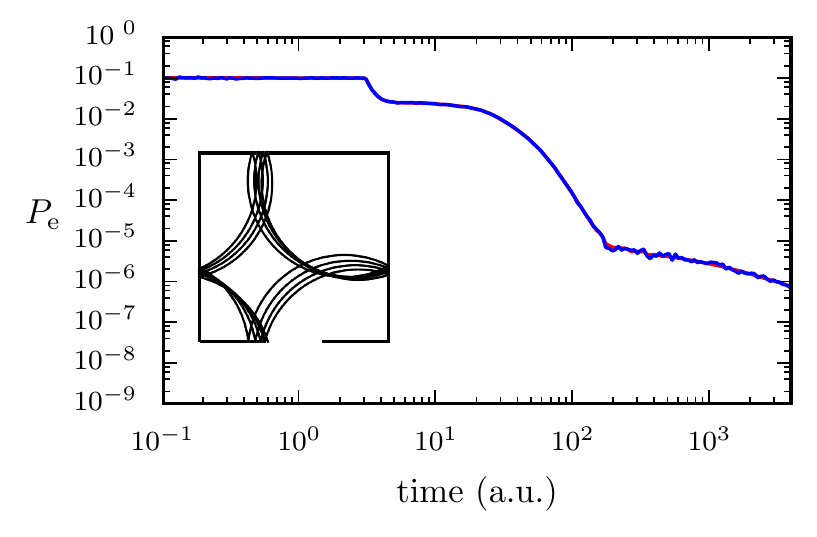}%
\caption{\label{fig:escape} Escape rate for a single-particle in square billiards in a magnetic field with Larmor radius $0.5075$. The result demonstrates a combination of algebraic and exponential behavior as a function of time. The escape rate computed from an
ensemble of trajectories using \texttt{bill2d\_escape} is shown as a blue line,
and a reference curve computed with an analytical collision map is shown as a
red line.}
\end{figure}

In Fig.~\ref{fig:escape} we show the escape rate for a
single particle square billiards in a magnetic field, as calculated with the
\texttt{bill2d\_escape} binary (blue line). The square has side length of unity, with a hole
of length~$0.3$, and the magnetic field is such that the Larmor radius is
$0.5075$. As a reference, we have calculated the escape rate using an
analytical mapping between boundary collisions (red line), which is available for such a
simple system. The results are nearly identical all the way to very small
escape rates (i.e., very rare trajectories). Note that trajectories
that can never escape the system are not included in the data.

After a brief initial period, the escape-rate curve is found to be a combination of exponential and algebraic functions, which is typical for systems with a mixed phase space in the corresponding closed system~\cite{leakingsystems}. In this system the mixed phase space (studied, e.g., in Ref.~\cite{berglundkunz}) is evident from, e.g., the regular characteristics of the example trajectory in the inset of Fig.~\ref{fig:escape}.

\subsection{Diffusion in periodic lattices}
When studying planar billiards in periodic structures, it is of interest to see how the particle diffuses -- on the average -- in the system. The diffusion coefficient is defined as
\begin{equation}
\label{eq:diffusioncoefficient}
D=\lim\limits_{t\to\infty} \frac{\langle \Vert \mathbf r(t) - \mathbf r(0) \Vert^2 \rangle}{4 t},
\end{equation}
where $\langle \cdot \rangle$ is the average over all possible initial configurations.

\begin{figure}[b!]\centering
\includegraphics[width=\linewidth]{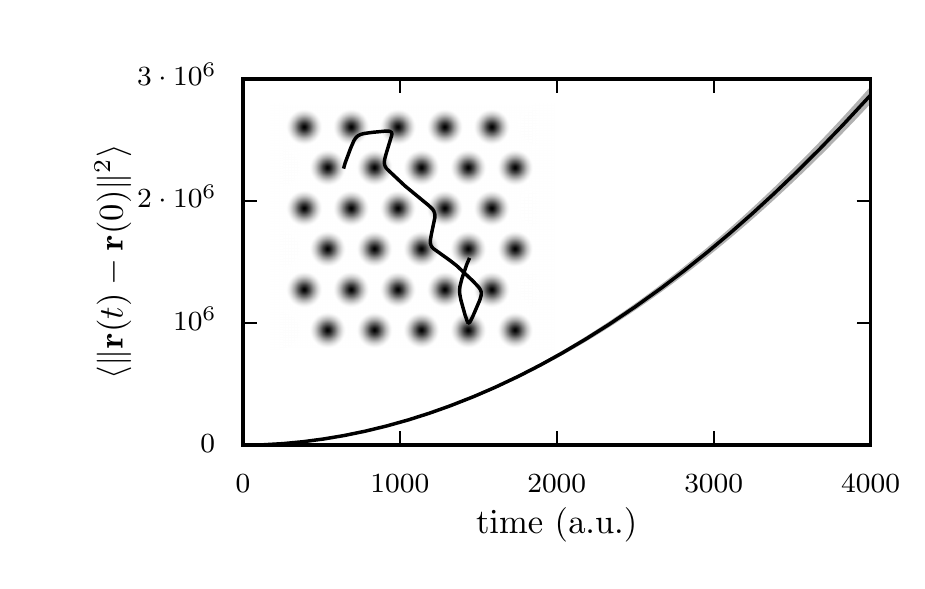}%
\caption{\label{fig:diffusion} Mean squared displacement for a soft Lorentz gas with scatterer radius $d=0.15$, softness $\sigma=0.1$, energy $E=\frac{1}{2}$ behaves as a power law in time ($\sim t^\gamma$) with exponent $\gamma\approx 2$. The inset shows a single trajectory in the system.}
\end{figure}

Let us consider a hexagonal antidot lattice of scatterers with lattice constant unity (see the inset of Fig.~\ref{fig:diffusion}).
Each scatterer produces a repulsive potential of the form $V_\alpha(\mathbf r) =
\left(1+\exp[(\Vert \mathbf r - \mathbf R_\alpha \Vert -
d)/\sigma]\right)^{-1}$. This is defined by three parameters: scatterer position $\mathbf R_\alpha$, radius $d$, and the softness of the potential $\sigma$.
The system thus corresponds to a Lorentz gas -- an extensively studied system in chaos theory -- with soft boundaries~\cite{lorentzreview}.

The mean squared displacement (MSD) $\langle \Vert \mathbf r(t) - \mathbf r(0)
\Vert^2 \rangle$ of a particle in the antidot lattice is shown in
Fig.~\ref{fig:diffusion} as a function of time. In this example the total
energy $E=\frac{1}{2}$, scatterer radius $d=0.15$, and scatterer softness
$\sigma=0.1$. The MSD has a power law behavior $\text{MSD} \sim t^{\gamma}$
with $\gamma\approx 2$, which means that the system is superdiffusive, and the
diffusion coefficient \eqref{eq:diffusioncoefficient} diverges. This behavior
is due to large number of trajectories demonstrating Lev\'y walk characteristics
or even purely ballistic behaviour.

In this example we have simulated an ensemble of $5500$ trajectories, which results in approx. 2.5 \% relative standard error of the mean for the MSD at $t=4000$. The calculation was done with the binary \emph{bill2d\_diffusion\_coefficient}.

\section{Summary}
\label{sec:summary}
The software package \textsc{Bill2d} is a versatile and efficient tool to simulate classical two-dimensional systems.
It supports, e.g., single and many-body simulations including varying particle number, interparticle
interactions, different external potentials, a magnetic field, periodic boundaries, and hard-wall billiard tables.
In addition to the versatility, the main strengths of \textsc{Bill2d} are its modular, object-oriented design, which allows
easy implementation of new features, and a clear and extensively documented source code.

\textsc{Bill2d} has already been used in several studies including, e.g., chaotic properties of many-body billiards~\cite{solanpaa_msc, circlebilliards},
nonlinear dynamics of Wigner molecules~\cite{wignercrystals}, and diffusion in periodic lattices~\cite{topithesis}.
The package can also be readily applied to, e.g., classical studies of transport properties in two-dimensional structures.

\section{Acknowledgments}
We are grateful for Stack Overflow user Rob$_\phi$ for his help with \texttt{Boost::program\_options},
Trond Norbye for his blog post on Python scripts and Automake, David Corvoysier for his blog post on GoogleTest and CMake,
Ryan Pavlik for his \texttt{GetGitRevisionDescription} CMake-module, Mark Moll for his post on CMake mailing list about CMake and Python,
and Visa Nummelin for initial development of the reference code for single-particle
magnetic rectangular billiards. We also thank Johannes Nokelainen, Sol Gil Gallegos, and Rainer Klages for useful discussions. Likewise, we thank
all our users for bug reports. This work was supported by the Academy of Finland
and the Finnish Cultural Foundation. We also acknowledge CSC -- the Finnish IT Center for Science for computational resources.

\bibliographystyle{elsarticle-num}

\begin{thebibliography}{10}
\expandafter\ifx\csname url\endcsname\relax
  \def\url#1{\texttt{#1}}\fi
\expandafter\ifx\csname urlprefix\endcsname\relax\def\urlprefix{URL }\fi
\expandafter\ifx\csname href\endcsname\relax
  \def\href#1#2{#2} \def\path#1{#1}\fi

\bibitem{openmushroom}
C.~P. Dettmann, O.~Georgiou, Open mushrooms: stickiness revisited, J. Phys. A
  44 (2011) 195102.
\newblock \href {http://dx.doi.org/10.1088/1751-8113/44/19/195102}
  {\path{doi:10.1088/1751-8113/44/19/195102}}.

\bibitem{newsinchaos}
L.~A. Bunimovich, L.~V. Vela-Arevalo, Some new surprises in chaos, Chaos 25
  (2015) 097614.
\newblock \href {http://dx.doi.org/10.1063/1.4916330}
  {\path{doi:10.1063/1.4916330}}.

\bibitem{circlebilliards}
J.~Solanp{\"a}{\"a}, J.~Nokelainen, P.~J.~J. Luukko, E.~R{\"a}s{\"a}nen,
  Coulomb-interacting billiards in circular cavities, J. Phys. A 46 (2013)
  235102.
\newblock \href {http://dx.doi.org/10.1088/1751-8113/46/23/235102}
  {\path{doi:10.1088/1751-8113/46/23/235102}}.

\bibitem{solanpaa_msc}
J.~Solanp\"a\"a,
  \href{http://www.solanpaa.fi/documents/solanpaa_msc_thesis.pdf}{Nonlinear
  dynamics and chaos in classical coulomb-interacting many-body billiards},
  Master's thesis, University of Jyv\"askyl\"a (2013).
\newline\urlprefix\url{http://www.solanpaa.fi/documents/solanpaa_msc_thesis.pdf}

\bibitem{openspherical}
C.~P. Dettmann, M.~R. Rahman, Survival probability for open spherical
  billiards, Chaos 24 (2014) 043130.
\newblock \href {http://dx.doi.org/10.1063/1.4900776}
  {\path{doi:10.1063/1.4900776}}.

\bibitem{stadiumtransport}
C.~P. Dettmann, O.~Georgiou, Transmission and reflection in the stadium
  billiard: Time-dependent asymmetric transport, Phys. Rev. E 83 (2011) 036212.
\newblock \href {http://dx.doi.org/10.1103/PhysRevE.83.036212}
  {\path{doi:10.1103/PhysRevE.83.036212}}.

\bibitem{bouncertransport}
C.~P. Dettmann, E.~D. Leonel, Escape and transport for an open bouncer:
  Stretched exponential decays, Physica D 241 (2012) 403 -- 408.
\newblock \href {http://dx.doi.org/10.1016/j.physd.2011.10.012}
  {\path{doi:10.1016/j.physd.2011.10.012}}.

\bibitem{escape_dettmann}
O.~Georgiou, C.~P. Dettmann, E.~G. Altmann, Faster than expected escape for a
  class of fully chaotic maps, Chaos 22 (2012) 043115.
\newblock \href {http://dx.doi.org/10.1063/1.4766723}
  {\path{doi:10.1063/1.4766723}}.

\bibitem{jpn_stickiness}
S.~Tsugawa, Y.~Aizawa, Stagnant motion in hamiltonian dynamics ---mushroom
  billiard case with smooth outermost kam tori---, J. Phys. Soc. Jpn., 83
  (2014) 024002.
\newblock \href {http://dx.doi.org/10.7566/JPSJ.83.024002}
  {\path{doi:10.7566/JPSJ.83.024002}}.

\bibitem{manyfaces}
L.~A. Bunimovich, L.~V. Vela-Arevalo, Many faces of stickiness in hamiltonian
  systems, Chaos 22 (2012) 026103.
\newblock \href {http://dx.doi.org/10.1063/1.3692974}
  {\path{doi:10.1063/1.3692974}}.

\bibitem{fine_stickiness}
L.~A. Bunimovich, Fine structure of sticky sets in mushroom billiards, J. Stat.
  Phys. 154 (2014) 421--431.
\newblock \href {http://dx.doi.org/10.1007/s10955-013-0898-2}
  {\path{doi:10.1007/s10955-013-0898-2}}.

\bibitem{lorentzreview}
C.~P. Dettmann, Diffusion in the lorentz gas, Commun. The. Phys. 62 (2014) 521.
\newblock \href {http://dx.doi.org/10.1088/0253-6102/62/4/10}
  {\path{doi:10.1088/0253-6102/62/4/10}}.

\bibitem{chaos_alive}
L.~A. Bunimovich, L.~V. Vela-Arevalo, Some new surprises in chaos, Chaos 25
  (2015) 097614.
\newblock \href {http://dx.doi.org/10.1063/1.4916330}
  {\path{doi:10.1063/1.4916330}}.

\bibitem{nakamura}
K.~Nakamura, T.~Harayama, Quantum Chaos and Quantum Dots, Oxford University
  Press, Oxford, 2003.

\bibitem{micolich}
A.~P. Micolich, A.~M. See, B.~C. Scannell, C.~A. Marlow, T.~P. Martin,
  I.~Pilgrim, A.~R. Hamilton, H.~Linke, R.~P. Taylor, Is it the boundaries or
  disorder that dominates electron transport in semiconductor `billiards'?,
  Fortschr. Physik 61 (2013) 332 -- 347.
\newblock \href {http://dx.doi.org/10.1002/prop.201200081}
  {\path{doi:10.1002/prop.201200081}}.

\bibitem{stadium_transport}
C.~P. Dettmann, O.~Georgiou, Transmission and reflection in the stadium
  billiard: Time-dependent asymmetric transport, Phys. Rev. E 83 (2011) 036212.
\newblock \href {http://dx.doi.org/10.1103/PhysRevE.83.036212}
  {\path{doi:10.1103/PhysRevE.83.036212}}.

\bibitem{alg_paper1}
M.~P. Allen, D.~J. Tildesley, Computer Simulation of Liquids, Oxford University
  Press (USA), 1989.
\newblock \href {http://dx.doi.org/10.1016/0167-7322(88)80022-9}
  {\path{doi:10.1016/0167-7322(88)80022-9}}.

\bibitem{alg_paper2}
R.~I. McLachlan, On the numerical integration of ordinary differential
  equations by symmetric composition methods, SIAM J. Sci. Comput. 16 (1995)
  151 -- 168.
\newblock \href {http://dx.doi.org/10.1137/0916010}
  {\path{doi:10.1137/0916010}}.

\bibitem{alg_paper3}
M.~Suzuki, General nonsymmetric higher-order decomposition of exponential
  operators and symplectic integrators, J. Phys. Soc. Jpn. 61 (1992) 3015 --
  3019.
\newblock \href {http://dx.doi.org/10.1143/JPSJ.61.3015}
  {\path{doi:10.1143/JPSJ.61.3015}}.

\bibitem{alg_paper4}
H.~Yoshida, Construction of higher order symplectic integrators, Phys. Lett. A
  150 (1990) 262 -- 268.
\newblock \href {http://dx.doi.org/10.1016/0375-9601(90)90092-3}
  {\path{doi:10.1016/0375-9601(90)90092-3}}.

\bibitem{spreiter-walter}
Q.~Spreiter, M.~Walter, Classical molecular dynamics simulation with the
  velocity verlet algorithm at strong external magnetic fields, J. Comp. Phys.
  152 (1999) 102 -- 119.
\newblock \href {http://dx.doi.org/10.1006/jcph.1999.6237}
  {\path{doi:10.1006/jcph.1999.6237}}.

\bibitem{4th_order_magfield_algorithm}
Y.~He, Y.~Sun, J.~Liu, H.~Qin, Volume-preserving algorithms for charged
  particle dynamics, J. Comput. Phys. 281 (2015) 135 -- 147.
\newblock \href {http://dx.doi.org/10.1016/j.jcp.2014.10.032}
  {\path{doi:10.1016/j.jcp.2014.10.032}}.

\bibitem{pre_77_066401}
S.~A. Chin, Symplectic and energy-conserving algorithms for solving magnetic
  field trajectories, Phys. Rev. E 77 (2008) 066401.
\newblock \href {http://dx.doi.org/10.1103/PhysRevE.77.066401}
  {\path{doi:10.1103/PhysRevE.77.066401}}.

\bibitem{c++standard}
{ISO/IEC} 14882:2014: {I}nformation technology -- {P}rogramming languages --
  {C}$++$, {I}nternational {O}rganization for {S}tandardization, {G}eneva,
  {S}witzerland, 2014.

\bibitem{mushroom}
L.~A. Bunimovich, Mushrooms and other billiards with divided phase space, Chaos
  11 (2001) 802 -- 808.
\newblock \href {http://dx.doi.org/10.1063/1.1418763}
  {\path{doi:10.1063/1.1418763}}.

\bibitem{K}
A.~N. Kolmogorov, On the conservation of conditionally periodic motions under
  small perturbation of the {H}amiltonian, Dokl. Akad. Nauk. SSSR 98 (1954) 527
  -- 320, in {R}ussian.

\bibitem{A}
V.~I. Arnold, Proof of a theorem of {A}. {N}. {K}olmogorov on the preservation
  of conditionally periodic motions under a small perturbation of the
  {H}amiltonian, Uspehi Mat. Nauk 18 (1963) 12 -- 40, in {R}ussian.

\bibitem{M}
J.~Moser, On invariant curves of area-preserving mappings of an annulus, Nachr.
  Akad. Wiss. Göttingen Math.-Phys. 2 (1962) 1 -- 20.

\bibitem{robnikberrymagfield}
M.~Robnik, M.~V. Berry, Classical billiards in magnetic fields, J. Phys. A 18
  (1985) 1361.
\newblock \href {http://dx.doi.org/10.1088/0305-4470/18/9/019}
  {\path{doi:10.1088/0305-4470/18/9/019}}.

\bibitem{leakingsystems}
E.~G. Altmann, J.~S.~E. Portela, T.~T\'el, Leaking chaotic systems, Rev. Mod.
  Phys. 85 (2013) 869 -- 918.
\newblock \href {http://dx.doi.org/10.1103/RevModPhys.85.869}
  {\path{doi:10.1103/RevModPhys.85.869}}.

\bibitem{berglundkunz}
N.~Berglund, H.~Kunz, Integrability and ergodicity of classical billiards in a
  magnetic field, J. Stat. Phys. 83 (1996) 81--126.
\newblock \href {http://dx.doi.org/10.1007/BF02183641}
  {\path{doi:10.1007/BF02183641}}.

\bibitem{wignercrystals}
J.~Solanp{\"a}{\"a}, P.~J.~J. Luukko, E.~R{\"a}s{\"a}nen, Many-particle
  dynamics and intershell effects in wigner molecules, J. Phys. Condens. Matter
  23 (2011) 395602.
\newblock \href {http://dx.doi.org/10.1088/0953-8984/23/39/395602}
  {\path{doi:10.1088/0953-8984/23/39/395602}}.

\bibitem{topithesis}
T.~H\"am\"al\"ainen, \href{http://URN.fi/URN:NBN:fi:tty-201405231225}{Diffusion
  calculations in the lorentz gas (english abstract)}, Master's thesis, Tampere
  University of Technology, Finland (2014).
\newline\urlprefix\url{http://URN.fi/URN:NBN:fi:tty-201405231225}

\end{thebibliography}


\begin{thebibliography}{10}
\expandafter\ifx\csname url\endcsname\relax
  \def\url#1{\texttt{#1}}\fi
\expandafter\ifx\csname urlprefix\endcsname\relax\def\urlprefix{URL }\fi
\expandafter\ifx\csname href\endcsname\relax
  \def\href#1#2{#2} \def\path#1{#1}\fi

\bibitem{chaos_alive}
L.~A. Bunimovich, L.~V. Vela-Arevalo, Some new surprises in chaos, Chaos 25
  (2015) 097614.
\newblock \href {http://dx.doi.org/10.1063/1.4916330}
  {\path{doi:10.1063/1.4916330}}.

\bibitem{circlebilliards}
J.~Solanp{\"a}{\"a}, J.~Nokelainen, P.~J.~J. Luukko, E.~R{\"a}s{\"a}nen,
  {Coulomb}-interacting billiards in circular cavities, J. Phys. A 46 (2013)
  235102.
\newblock \href {http://dx.doi.org/10.1088/1751-8113/46/23/235102}
  {\path{doi:10.1088/1751-8113/46/23/235102}}.

\bibitem{solanpaa_msc}
J.~Solanp\"a\"a,
  \href{http://www.solanpaa.fi/documents/solanpaa_msc_thesis.pdf}{Nonlinear
  dynamics and chaos in classical {Coulomb}-interacting many-body billiards},
  Master's thesis, University of Jyv\"askyl\"a (2013).
\newline\urlprefix\url{http://www.solanpaa.fi/documents/solanpaa_msc_thesis.pdf}

\bibitem{openspherical}
C.~P. Dettmann, M.~R. Rahman, Survival probability for open spherical
  billiards, Chaos 24 (2014) 043130.
\newblock \href {http://dx.doi.org/10.1063/1.4900776}
  {\path{doi:10.1063/1.4900776}}.

\bibitem{stadium_transport}
C.~P. Dettmann, O.~Georgiou, Transmission and reflection in the stadium
  billiard: Time-dependent asymmetric transport, Phys. Rev. E 83 (2011) 036212.
\newblock \href {http://dx.doi.org/10.1103/PhysRevE.83.036212}
  {\path{doi:10.1103/PhysRevE.83.036212}}.

\bibitem{universal_decays}
G.~Cristadoro, R.~Ketzmerick, Universality of algebraic decays in hamiltonian
  systems, Phys. Rev. Lett. 100 (2008) 184101.
\newblock \href {http://dx.doi.org/10.1103/PhysRevLett.100.184101}
  {\path{doi:10.1103/PhysRevLett.100.184101}}.

\bibitem{bouncertransport}
C.~P. Dettmann, E.~D. Leonel, Escape and transport for an open bouncer:
  Stretched exponential decays, Physica D 241 (2012) 403 -- 408.
\newblock \href {http://dx.doi.org/10.1016/j.physd.2011.10.012}
  {\path{doi:10.1016/j.physd.2011.10.012}}.

\bibitem{round_border_escape}
M.~S. Cust\'odio, M.~W. Beims, Intrinsic stickiness and chaos in open
  integrable billiards: Tiny border effects, Phys. Rev. E 83 (2011) 056201.
\newblock \href {http://dx.doi.org/10.1103/PhysRevE.83.056201}
  {\path{doi:10.1103/PhysRevE.83.056201}}.

\bibitem{escape_dettmann}
O.~Georgiou, C.~P. Dettmann, E.~G. Altmann, Faster than expected escape for a
  class of fully chaotic maps, Chaos 22 (2012) 043115.
\newblock \href {http://dx.doi.org/10.1063/1.4766723}
  {\path{doi:10.1063/1.4766723}}.

\bibitem{openmushroom}
C.~P. Dettmann, O.~Georgiou, Open mushrooms: stickiness revisited, J. Phys. A
  44 (2011) 195102.
\newblock \href {http://dx.doi.org/10.1088/1751-8113/44/19/195102}
  {\path{doi:10.1088/1751-8113/44/19/195102}}.

\bibitem{jpn_stickiness}
S.~Tsugawa, Y.~Aizawa, Stagnant motion in {Hamiltonian} dynamics ---mushroom
  billiard case with smooth outermost {KAM} tori---, J. Phys. Soc. Jpn., 83
  (2014) 024002.
\newblock \href {http://dx.doi.org/10.7566/JPSJ.83.024002}
  {\path{doi:10.7566/JPSJ.83.024002}}.

\bibitem{manyfaces}
L.~A. Bunimovich, L.~V. Vela-Arevalo, Many faces of stickiness in {Hamiltonian}
  systems, Chaos 22 (2012) 026103.
\newblock \href {http://dx.doi.org/10.1063/1.3692974}
  {\path{doi:10.1063/1.3692974}}.

\bibitem{fine_stickiness}
L.~A. Bunimovich, Fine structure of sticky sets in mushroom billiards, J. Stat.
  Phys. 154 (2014) 421--431.
\newblock \href {http://dx.doi.org/10.1007/s10955-013-0898-2}
  {\path{doi:10.1007/s10955-013-0898-2}}.

\bibitem{lorentzreview}
C.~P. Dettmann, Diffusion in the {Lorentz} gas, Commun. The. Phys. 62 (2014)
  521.
\newblock \href {http://dx.doi.org/10.1088/0253-6102/62/4/10}
  {\path{doi:10.1088/0253-6102/62/4/10}}.

\bibitem{flower_lorentz}
T.~Harayama, R.~Klages, P.~Gaspard, Deterministic diffusion in flower-shaped
  billiards, Phys. Rev. E 66 (2002) 026211.
\newblock \href {http://dx.doi.org/10.1103/PhysRevE.66.026211}
  {\path{doi:10.1103/PhysRevE.66.026211}}.

\bibitem{3dlorentz}
T.~Gilbert, H.~C. Nguyen, D.~P. Sanders, Diffusive properties of persistent
  walks on cubic lattices with application to periodic {Lorentz} gases, Journal
  of Physics A: Mathematical and Theoretical 44 (2011) 065001.
\newblock \href {http://dx.doi.org/10.1088/1751-8113/44/6/065001}
  {\path{doi:10.1088/1751-8113/44/6/065001}}.

\bibitem{recent_hard_wall_lorentz}
G.~Cristadoro, T.~Gilbert, M.~Lenci, D.~P. Sanders, Measuring logarithmic
  corrections to normal diffusion in infinite-horizon billiards, Phys. Rev. E
  90 (2014) 022106.
\newblock \href {http://dx.doi.org/10.1103/PhysRevE.90.022106}
  {\path{doi:10.1103/PhysRevE.90.022106}}.

\bibitem{magnetoresistance}
J.~Wiersig, K.-H. Ahn, Devil's staircase in the magnetoresistance of a periodic
  array of scatterers, Phys. Rev. Lett. 87 (2001) 026803.
\newblock \href {http://dx.doi.org/10.1103/PhysRevLett.87.026803}
  {\path{doi:10.1103/PhysRevLett.87.026803}}.

\bibitem{magnetoresistance2}
M.~Khoury, A.~M. Lacasta, J.~M. Sancho, A.~H. Romero, K.~Lindenberg, Charged
  particle transport in antidot lattices in the presence of magnetic and
  electric fields: Langevin approach, Phys. Rev. B 78 (2008) 155433.
\newblock \href {http://dx.doi.org/10.1103/PhysRevB.78.155433}
  {\path{doi:10.1103/PhysRevB.78.155433}}.

\bibitem{diffusion_repulsive_attractive}
J.~Yang, H.~Zhao, Anomalous diffusion among two-dimensional scatterers, Journal
  of Statistical Mechanics: Theory and Experiment 2010 (2010) L12001.
\newblock \href {http://dx.doi.org/10.1088/1742-5468/2010/12/L12001}
  {\path{doi:10.1088/1742-5468/2010/12/L12001}}.

\bibitem{diffusion_2d_soft}
N.-C. Panoiu, Anomalous diffusion in two-dimensional potentials with hexagonal
  symmetry, Chaos 10 (2000) 166--179.
\newblock \href {http://dx.doi.org/http://dx.doi.org/10.1063/1.166484}
  {\path{doi:http://dx.doi.org/10.1063/1.166484}}.

\bibitem{nakamura}
K.~Nakamura, T.~Harayama, Quantum Chaos and Quantum Dots, Oxford University
  Press, Oxford, 2003.

\bibitem{micolich}
A.~P. Micolich, A.~M. See, B.~C. Scannell, C.~A. Marlow, T.~P. Martin,
  I.~Pilgrim, A.~R. Hamilton, H.~Linke, R.~P. Taylor, Is it the boundaries or
  disorder that dominates electron transport in semiconductor `billiards'?,
  Fortschr. Physik 61 (2013) 332 -- 347.
\newblock \href {http://dx.doi.org/10.1002/prop.201200081}
  {\path{doi:10.1002/prop.201200081}}.

\bibitem{alg_paper1}
M.~P. Allen, D.~J. Tildesley, Computer Simulation of Liquids, Oxford University
  Press (USA), 1989.
\newblock \href {http://dx.doi.org/10.1016/0167-7322(88)80022-9}
  {\path{doi:10.1016/0167-7322(88)80022-9}}.

\bibitem{alg_paper2}
R.~I. McLachlan, On the numerical integration of ordinary differential
  equations by symmetric composition methods, SIAM J. Sci. Comput. 16 (1995)
  151 -- 168.
\newblock \href {http://dx.doi.org/10.1137/0916010}
  {\path{doi:10.1137/0916010}}.

\bibitem{alg_paper3}
M.~Suzuki, General nonsymmetric higher-order decomposition of exponential
  operators and symplectic integrators, J. Phys. Soc. Jpn. 61 (1992) 3015 --
  3019.
\newblock \href {http://dx.doi.org/10.1143/JPSJ.61.3015}
  {\path{doi:10.1143/JPSJ.61.3015}}.

\bibitem{alg_paper4}
H.~Yoshida, Construction of higher order symplectic integrators, Phys. Lett. A
  150 (1990) 262 -- 268.
\newblock \href {http://dx.doi.org/10.1016/0375-9601(90)90092-3}
  {\path{doi:10.1016/0375-9601(90)90092-3}}.

\bibitem{spreiter-walter}
Q.~Spreiter, M.~Walter, Classical molecular dynamics simulation with the
  velocity {Verlet} algorithm at strong external magnetic fields, J. Comp.
  Phys. 152 (1999) 102 -- 119.
\newblock \href {http://dx.doi.org/10.1006/jcph.1999.6237}
  {\path{doi:10.1006/jcph.1999.6237}}.

\bibitem{4th_order_magfield_algorithm}
Y.~He, Y.~Sun, J.~Liu, H.~Qin, Volume-preserving algorithms for charged
  particle dynamics, J. Comput. Phys. 281 (2015) 135 -- 147.
\newblock \href {http://dx.doi.org/10.1016/j.jcp.2014.10.032}
  {\path{doi:10.1016/j.jcp.2014.10.032}}.

\bibitem{pre_77_066401}
S.~A. Chin, Symplectic and energy-conserving algorithms for solving magnetic
  field trajectories, Phys. Rev. E 77 (2008) 066401.
\newblock \href {http://dx.doi.org/10.1103/PhysRevE.77.066401}
  {\path{doi:10.1103/PhysRevE.77.066401}}.

\bibitem{c++standard}
{ISO/IEC} 14882:2014: {I}nformation technology -- {P}rogramming languages --
  {C}$++$, {I}nternational {O}rganization for {S}tandardization, {G}eneva,
  {S}witzerland, 2014.

\bibitem{mushroom}
L.~A. Bunimovich, Mushrooms and other billiards with divided phase space, Chaos
  11 (2001) 802 -- 808.
\newblock \href {http://dx.doi.org/10.1063/1.1418763}
  {\path{doi:10.1063/1.1418763}}.

\bibitem{K}
A.~N. Kolmogorov, On the conservation of conditionally periodic motions under
  small perturbation of the {H}amiltonian, Dokl. Akad. Nauk. SSSR 98 (1954) 527
  -- 320, in {R}ussian.

\bibitem{A}
V.~I. Arnold, Proof of a theorem of {A}. {N}. {K}olmogorov on the preservation
  of conditionally periodic motions under a small perturbation of the
  {H}amiltonian, Uspehi Mat. Nauk 18 (1963) 12 -- 40, in {R}ussian.

\bibitem{M}
J.~Moser, On invariant curves of area-preserving mappings of an annulus, Nachr.
  Akad. Wiss. Göttingen Math.-Phys. 2 (1962) 1 -- 20.

\bibitem{robnikberrymagfield}
M.~Robnik, M.~V. Berry, Classical billiards in magnetic fields, J. Phys. A 18
  (1985) 1361.
\newblock \href {http://dx.doi.org/10.1088/0305-4470/18/9/019}
  {\path{doi:10.1088/0305-4470/18/9/019}}.

\bibitem{leakingsystems}
E.~G. Altmann, J.~S.~E. Portela, T.~T\'el, Leaking chaotic systems, Rev. Mod.
  Phys. 85 (2013) 869 -- 918.
\newblock \href {http://dx.doi.org/10.1103/RevModPhys.85.869}
  {\path{doi:10.1103/RevModPhys.85.869}}.

\bibitem{berglundkunz}
N.~Berglund, H.~Kunz, Integrability and ergodicity of classical billiards in a
  magnetic field, J. Stat. Phys. 83 (1996) 81--126.
\newblock \href {http://dx.doi.org/10.1007/BF02183641}
  {\path{doi:10.1007/BF02183641}}.

\bibitem{wignercrystals}
J.~Solanp{\"a}{\"a}, P.~J.~J. Luukko, E.~R{\"a}s{\"a}nen, Many-particle
  dynamics and intershell effects in {Wigner} molecules, J. Phys. Condens.
  Matter 23 (2011) 395602.
\newblock \href {http://dx.doi.org/10.1088/0953-8984/23/39/395602}
  {\path{doi:10.1088/0953-8984/23/39/395602}}.

\bibitem{topithesis}
T.~H\"am\"al\"ainen, \href{http://URN.fi/URN:NBN:fi:tty-201405231225}{Diffusion
  calculations in the {Lorentz} gas (english abstract)}, Master's thesis,
  Tampere University of Technology, Finland (2014).
\newline\urlprefix\url{http://URN.fi/URN:NBN:fi:tty-201405231225}

\end{thebibliography}

\end{document}